\documentclass[prb,twocolumn,aps,showpacs,fixfloats]{revtex4-1}
\usepackage{graphicx}
\usepackage{bm}
\usepackage{amsmath,amssymb}
\usepackage{subfigure}
\usepackage{float}
\usepackage{latexsym}
\usepackage{epstopdf}
\usepackage{color}
\usepackage{pdfpages}

\newcommand{\s}{\scriptscriptstyle}

\begin{document}

\title {Effective spin Hall properties of a mixture of materials with and without spin-orbit coupling: Tailoring the effective spin-diffusion length}

\author{Z. Yue$^{1}$, M. C. Prestgard$^{2}$, A. Tiwari$^{2}$, and M. E. Raikh$^{1}$ }

\affiliation{$^{1}$Department of Physics and
Astronomy, University of Utah, Salt Lake City, UT 84112, USA \\
$^{2}$Department of Materials Science and Engineering, University of Utah, Salt Lake City, Utah 84112, USA
}

\begin{abstract}
We study theoretically the effective spin Hall properties of a composite consisting
of  two materials with and without spin-orbit (SO) coupling. In particular, we assume
that SO material represents a system of grains in a matrix with no SO. We calculate the
 effective spin Hall angle and the effective spin diffusion length of the mixture.
Our main qualitative finding is that, when the bare spin diffusion length is much smaller than the
radius of the grain, the {\em effective} spin diffusion length is strongly enhanced, well beyond
the ``geometrical" factor. The physical origin of this additional enhancement is that, with small diffusion length, the spin current mostly flows {\em around the grain}  without suffering
much loss. We also demonstrate that the voltage, created by a spin current, is sensitive to a very weak magnetic field  directed along the spin current, and even reverses sign in a certain domain of fields. The origin of this sensitivity is that the spin precession, caused by magnetic field, takes place outside the grains where SO is absent.

\end{abstract}

\maketitle

\section{Introduction}

%Definition of the spin Hall angle: $j_{\s c} =\theta_{\s \text{SH}}(2e/\hbar)j_{\s s}$

The spin Hall effect\cite{Dyakonov1,Dyakonov2,Hirsch}(SHE), predicted theoretically more than four decades ago\cite{Dyakonov1,Dyakonov2}, is nowadays routinely observed in many
materials,\cite {Pt1,Pt2,Pt3,Pt4,Pt5,Pt6,Bader,Gold,Exotic,GaAs,silicon1,silicon2,germanium,graphene}
which include traditional and exotic metals, prominent semiconductors, and graphene.
Moreover, the inverse spin Hall effect (ISHE), i.e. generation of voltage drop normal to the spin current, was recently ``put to work".
It serves as a tool to detect whether or not the spin current
is injected into a nonmagnetic material from an ac-driven ferromagnet in the course of spin pumping.
Most recently\cite{Polaron1,Polaron2,Polaron3,Polaron4} the pumped spin currents in certain polymers were registered via inverse spin Hall voltage which they induced in  Pt electrode located at some distance from the interface with ferromagnet.

\begin{figure}
\includegraphics[width=70mm]{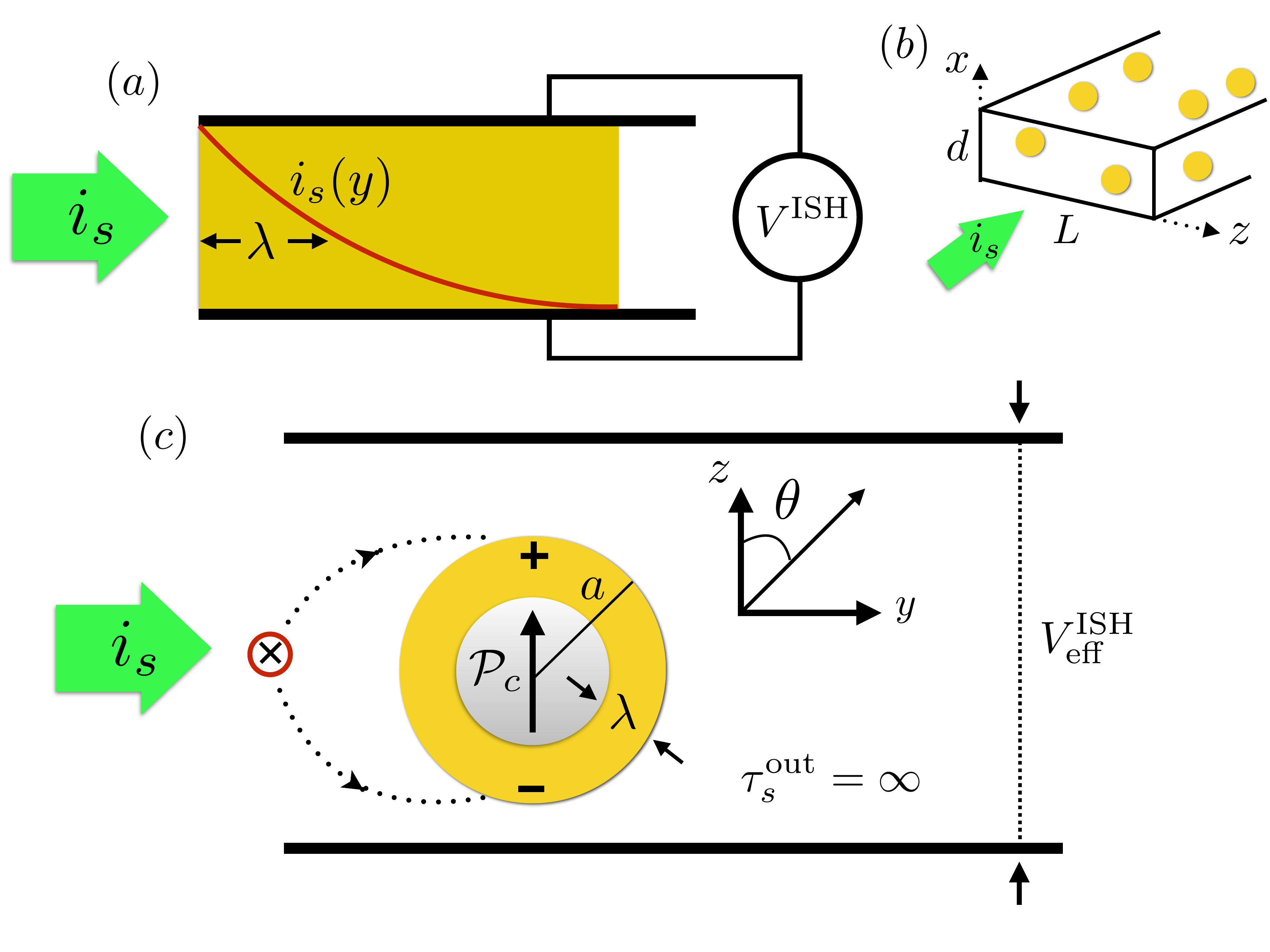}
\caption{(Color online) (a) Conventional geometry for the inverse spin Hall effect. Spin current flowing along $y$ causes a buildup of the voltage, $V^{\s \text{ISH}}$, between the edges $z=\pm L/2$. The buildup takes place as long as $y$ is smaller than the spin diffusion length, $\lambda$.  (b) Schematic illustration of a ``granular" geometry, where
the SO-coupled material is dissolved in the matrix with no SO coupling.
(c) microscopic scenario of ISHE on a single spherical granule of a radius, $a$. Spin
current with polarization along $x$ turns the sphere into an electrical dipole directed normally to the current. The magnitude of a dipole moment, ${\cal P}_{\s c}$, depends on the ratio between $a$ and $\lambda$, while the electric field inside the granule is homogeneous.}
\label{Fig1}
\end{figure}

The latest focus\cite{EitanWithPt,VardenyWithPt,BoehmeWithPt} of the research on the spin physics in organics  is the study of the properties of
platinum-containing $\pi$-conjugated polymers. In these materials Pt atoms are embedded in the polymer backbone chains.  While the SO coupling, which is the origin of the  SHE, is very weak in polymers, adding of Pt creates the elements of the backbone
 where it is locally strong. These elements can be separated either  by one or by three $\pi$-conjugated
spacer unit lengths. In this regard, a general question arises: how the spin Hall effect is realized   in composite materials where the strong SO and low SO domains are intermixed?
Note that, by now,   all theoretical studies of SO-related  transport assumed that the SO coupling is homogeneous.

The goal of the present paper is to develop an elementary theory which addresses the question formulated above. Unlike Refs. \onlinecite{Yu1,Yu2}, we will not
specify a mechanism of SO on the microscopic level, but rather focus on purely ``geometrical" aspects. Namely, we will consider the following minimal model: a system of SO grains is dissolved in a matrix with no SO. The question we will be interested in is: what are the effective spin Hall characteristics of the mixture.

%{\bf If SO molecules are incorporated into a polymer\cite{EitanWithPt,VardenyWithPt,BoehmeWithPt}, they occupy only a small porion of the volume. The current mostly bypasses them. The the question arises: whether these molecules would
%cause the voltage buildup at the edges of the sample, and, if yes, what is the magnitude of this
%voltage? Also, since the spin relaxation occurs predominantly inside the molecules, where SO is strong, what is the effective spin relaxation rate? Unlike Refs. \onlinecite{Yu1,Yu2} we will not
%specify the mechanism of SO on the microscopic level, but rather focus on purely ``geometrical" aspects. Namely, we will consider the following minimal model: a system of SO grains dissolved in a matrix with no SO. The questions of interest are the effective spin Hall properties of the mixture.}
 Firstly, we address a mechanism of the formation of the inverse spin Hall voltage between the
 edges of the sample in the geometry of the mixture. Unlike the case of homogeneous SO, this formation happens as follows. The spin current turns each SO grain into an electric dipole.
 All dipole moments are oriented normal to the spin current. Thus the potentials they create at the
upper and the lower boundaries of the sample add up. The difference of these potentials is the effective ISH voltage, $V^{\s \text{SH}}_{\s \text{eff}}$, of the mixture, which can be related
to the effective spin Hall angle, $\theta^{\s \text{SH}}_{\s \text{eff}}$.

Naively, one would expect that, in a mixture of grains of density, $n$, and radius, $a$,
the relation  $\theta^{\s \text{SH}}_{\s \text{eff}}=(na^3)\theta^{\s \text{SH}}$ holds within a numerical factor. Here  $\theta^{\s \text{SH}}$ is the spin Hall angle of the bulk SO material. This is simply because $na^3$ is the volume fraction of the SO material.
Equally, one would  expect that the effective spin relaxation time of the mixture is $1/(na^3)$ times
longer than in the SO material, so that
spin diffusion length,  $\lambda_{\s \text{eff}}$  is related to the spin diffusion length, $\lambda$, of the SO material as
$\lambda_{\s \text{eff}}=(na^3)^{-1/2}\lambda$.

The above expectations are correct only in the limit when the grains are small enough,
namely, $a \ll \lambda$, so that the portion of spin polarization, which is lost within a single grain, is small. The opposite case of large grains, $a \gg \lambda$, is much less trivial. As we show below,
in this limit $V^{\s \text{SH}}_{\s \text{eff}}\sim \lambda na^2 V^{\s \text{SH}}$, while
$\lambda_{\s \text{eff}} \sim \frac{1}{(na)^{1/2}}$. In other words, at small  $\lambda$,  the effective spin-diffusion length {\em saturates}. This finding can be loosely interpreted from the perspective of
 diffusion  in the presence of the absorbing traps. The stronger is the absorption, the smaller is the concentration of particles at the position of the trap.

Finally, we will demonstrate that $V^{\s \text{SH}}_{\s \text{eff}}$ is sensitive to a very weak magnetic field.
In a homogeneous material, the spin Hall effect gets suppressed in the field with Larmour frequency $\Omega \sim \tau_{\s s}^{-1}$, where $\tau_{\s s}$ is the spin-relaxation time. For the mixture, the characteristic field is $\sim T^{-1}$, where $T$ is the diffusion time {\em between the sample edges}. This is because  spin precession takes place mostly outside the grains.
The paper is organized as follows. In Sect. II we solve an auxiliary problem of electric the polarization of a grain with a given radius, $a$, by the spin current. The solution is then employed to calculate the
effective inverse spin Hall voltage in the mixture of grains with concentration, $n$. Sensitivity of this voltage to a weak longitudinal  magnetic field is studied in Sect. III. In Sect. IV the effective diffusion length, $\lambda_{\s \text{eff}}$, of the mixture is expressed via $\lambda$,  $a$, and the parameter $na^3$. The physics of elongation of $\lambda_{\s \text{eff}}$ for small $\lambda \ll a$ is discussed in Sect.~V. Concluding remarks are presented in Sect. VI.

\section{Calculation of effective characteristics of the mixture}
\subsection{Single grain}
\begin{figure}
\includegraphics[width=70mm]{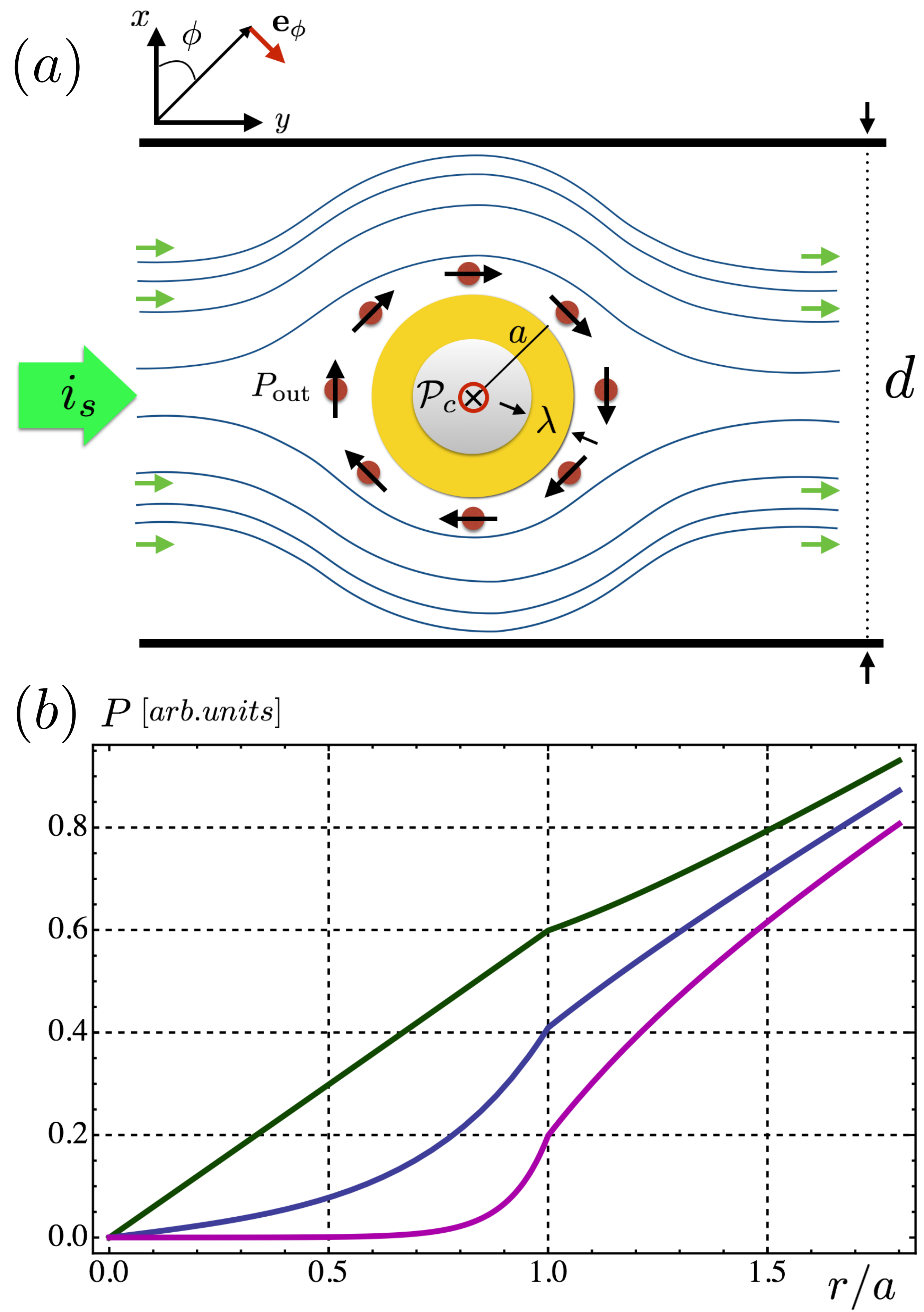}
\caption{(Color online) (a) The cross section $z=0$. Distribution of the  spin current
in the $(x,y)$-plane in the presence of a spherical grain, Eq. (\ref{35}), is illustrated schematically for $r>a$. Inside the grain, $r<a$, this distribution is determined by Eq. (\ref{27}). (b) Distribution of the spin polarization along the radius, $r$, is plotted for
$D_{\s \text{out}}/D_{\s \text{in}}=2$ and three values of $\lambda$: $a/\lambda =0.2$ (green),
$a/\lambda =4$ (blue), and $a/\lambda =12$ (purple). Enhancement of the effective spin
diffusion length for small $\lambda/a$ is a result of a strong suppression of polarization
near the boundary $r=a$.}
\label{Fig2}
\end{figure}

 The simplest way to incorporate the spin Hall effect on a quantitative level\cite{DyakonovMagnetoresistance}
is to add to the current density, ${\bm j}=\sigma {\bm E}$,  the term
%$\gamma D\nabla \times{\bm P}$,
$\gamma D\hspace{0.5mm}\text{curl}\hspace{0.5mm}{\bm P}$
where $\sigma$ and $D$ are the conductivity and   the diffusion coefficient, respectively,
${\bm P}({\bm r})$ is the coordinate-dependent spin polarization. The strength of the SO coupling
is quantified by a dimensionless parameter $\gamma$.
The system of coupled equations for the spatial distribution of ${\bm P}({\bm r})$ and ${\bm j}({\bm r})$ reads \cite{DyakonovMagnetoresistance}
\begin{align}
\label{1}
{\bm j}&=\sigma{\bm E}+e\gamma D\hspace{0.5mm}\text{curl}\hspace{0.5mm}{\bm P}. \\
\label{2}
q_{\s ij}&=-D\frac{\partial P_{\s j}}{\partial x_{\s i}}+\frac{\gamma}{e} \sigma \varepsilon_{\s ijk} E_{\s k}.
\end{align}
The second equation defines the component $i$ of the flux of the $j$-projection of spin polarization. The system becomes closed\cite{DyakonovMagnetoresistance} when it is complemented by the continuity equation
\begin{equation}
\label{3}
%\frac{\partial P_{\s j}}{\partial t}+
\frac{\partial q_{\s ij}}{\partial x_{\s i}}+\frac{ P_{\s j}}{\tau_{\s s}}=0.
\end{equation}

Consider an isolated spherical grain with radius, $a$, and with  the strength of SO-coupling, $\gamma$, embedded into an infinite medium with $\gamma=0$ and with no spin relaxation, $\tau_{\s s}=\infty$,
 Fig. \ref{Fig1}. Assume that the flux of
spins, oriented along the $x$-axis and  flowing along  the $y$-axis, is incident on the grain.
In application to the geometry, Fig. \ref{Fig1}, the essence of the inverse spin Hall effect is that the incident spin current, $i_{\s s}$, induces an effective  electric dipole on the sphere.
The induced dipole moment is perpendicular to both, the current direction and polarization direction in the incident flux, i.e. it is directed along the $z$-axis.

To calculate the magnitude, ${\cal P}_{\s c}$, of the dipole moment
it is natural to switch to spherical coordinates in which
the incident polarization, $P_{\s x}=-\frac{i_{\s s}}{D_{\s \text{out}}}y$, and the spin-current
density, $i_{\s y}=i_{\s s}$, have the form
\begin{equation}
\label{4}
{\bm P}=-\frac{i_{\s s}}{D_{\s \text{out}}}r\sin\theta \hspace{0.5mm}{\bm e_{\s \phi}},~~~~{\bm i}_{\s s}=i_{\s s}(\sin\theta \hspace{0.5mm} {\bm e_{\s r}}+\cos \theta \hspace{0.5mm} {\bm  e_{\s \theta}}),
\end{equation}
where ${\bm e_{\s r}}$, ${\bm e_{\s \theta}}$, and ${\bm e_{\s \phi}}$ are the unit vectors along
radial, polar, and azimuthal axes, respectively, see Fig. \ref{Fig2}.

Induced dipole moment along $z$ creates an electrostatic potential,
\begin{equation}
\label{5}
\varphi_{\s \text{out}} =\frac{{\cal P}_{\s c}\cos \theta}{r^2},
\end{equation}
outside the sphere.

From the form of $\varphi_{\s \text{out}}$ we conclude that the $\theta$-dependence of $\varphi$ inside the sphere is also proportional to $\cos \theta$.
%$\varphi\propto \cos\theta$.
This, together with Poisson's equation $\Delta \varphi =0$,
suggests that the induced electric field, ${\bm E}_{\s \text{in}}$, inside the sphere is homogeneous, so that
\begin{equation}
\label{6}
\varphi_{\s \text{in}}=-E_{\s \text{in}}r\cos\theta.
\end{equation}

Substituting Eq. (\ref{2}) into Eq.(\ref{3}), and taking into account that $\partial{\bm E}_{\s \text{in}}/\partial x_{\s i}=0$,
%With ${\bm E}_{\s \text{in}}$ being homogeneous,
we conclude that all the components of polarization inside the sphere satisfy the diffusion equation
\begin{equation}
\label{7}
D_{\s \text{in}}\Delta P_j+\frac{P_j}{\tau_{\s s}}=0,
\end{equation}
where $D_{\s \text{in}}$ is the diffusion coefficient inside the sphere.

As we will see below, the polarization, ${\bm P}({\bm r})$,
has only $\phi$-component inside the sphere and at all distances
outside the sphere.  As in the incident flux, Eq. (\ref{4}),
the angular dependence of $P_{\phi}$ is
$\propto \sin\theta$. Outside the sphere, where $\Delta {\bm P}=0$,
the general form of  $P_{\phi}$ is
\begin{equation}
\label{8}
{\bm P}_{\s \text{out}}=-\frac{i_{\s s}}{D_{\s \text{out}}}(r+\frac{\chi_{\s s}}{r^2})\sin\theta \hspace{0.5mm}{\bm e_{\s \phi}},
\end{equation}
where the constant $\chi_{\s s}$ is the ``spin polarizability".
Inside the sphere, the solution of Eq. (\ref{7}), proportional to $\sin \theta$, has the form
\begin{equation}
\label{9}
{\bm P}_{\s \text{in}}={\tilde P}~ i_1 (r/\lambda)\sin\theta \hspace{0.5mm}{\bm e}_{\s \phi},
\end{equation}
where ${\tilde P}$ is a constant, and
\begin{equation}
\label{9a}
\lambda=\left(D_{\s \text{in}} \tau_{\s s}\right)^{1/2} ,
\end{equation}
is the diffusion length. The function  $i_1(x)$ is a modified spherical Bessel function. We chose the function $i_1$ because it is finite at $x=0$.

While the polarization has only $\phi$-component, the spin current, defined as a flow of the $\phi$-component of spin, can be presented in the vector form
\begin{equation}
\label{9b}
{\bm i}_{\s \phi}=i_{\s s}\Big[(1-\frac{2\chi_{\s s}}{r^3})\sin\theta{\bm e}_{\s r}+
(1+\frac{\chi_{\s s}}{r^3})\cos\theta{\bm e}_{\s \theta}\Big],
\end{equation}
where the first term is $\partial P^{\s \phi}_{\s \text{out}}/\partial r$, while the second term is $\left(\frac{1}{r}\right)\partial P^{\s \phi}_{\s \text{out}}/\partial \theta$.
%We remind that the
At large distances the current Eq. (\ref{9b}) reproduces Eq. (\ref{4}).
%flows in the medium with no SO.

There are two unknown constants, ${\cal P}_{\s c}$ and $\chi_{\s s}$, in the expressions for electric field and spin polarization
inside the sphere, and two unknown constants, $E_{\s in}$ and ${\tilde P}$, in the corresponding expressions outside the sphere. These constants are determined from the four
boundary conditions at $r=a$:

(i)  Continuity of the tangent component of electric field
\begin{equation}
\label{10}
E_{\s \text{in}}=-\frac{{\cal P}_{\s c}}{a^3}.
\end{equation}

(ii) Continuity of the normal component of the charge current
\begin{equation}
\label{11}
\sigma_{\s \text{in}} E_{\s \text{in}}+\frac{2e\gamma D_{\s \text{in}}{\tilde P}}{
a}i_1(a/\lambda)=\frac{2\sigma_{\s \text{out}}{\cal P}_{\s c}}{a^3}.
\end{equation}

(iii) Continuity of the spin polarization
\begin{equation}
\label{12}
-\frac{i_{\s s}}{D_{\s \text{out}}}\Big(a+\frac{\chi_{\s s}}{a^2}\Big) ={\tilde P}i_1(a/\lambda).
\end{equation}

(iiii) Continuity of the spin flux though the boundary
\begin{equation}
\label{13}
\frac{D_{\s \text{in}}{\tilde P}}{\lambda}i'_1(a/\lambda)+\frac{\gamma}{e} \sigma_{\s \text{in}}E_{\s \text{in}}=i_{\s s}\Big(\frac{2\chi_{\s s}}{a^3}-1\Big).
\end{equation}

The system Eqs. (\ref{10})-(\ref{13}) yields the sought expression for the spin-current-induced dipole moment
\begin{equation}
\label{14}
{\cal P}_{\s c}=-\frac{6e a^3\gamma }{(\sigma_{\s \text{in}}+2\sigma_{\s \text{out}} ){\cal M}}i_{\s s},
\end{equation}
where ${\cal M}$ in denominator is the dimensionless combination
\begin{equation}
\label{15}
{\cal M}=\frac{2D_{\s \text{out}}}{D_{\s \text{in}}}-\frac{2\gamma^2\sigma_{\s \text{in}}}{\sigma_{\s \text{in}}+2\sigma_{\s \text{out}}}+\frac{a i'_1(\frac{a}{\lambda})}{\lambda i_1(\frac{a}{\lambda})}
\end{equation}
Naturally, the proportionality coefficient between ${\cal P}_{\s c}$ and the spin current contains the first power of the SO coupling strength, $\gamma$.

The second term in Eq. (\ref{15}) contains $\gamma^2$, and can be safely neglected. The ratio
$D_{\s \text{out}}/D_{\s \text{in}}$ can be replaced by $\sigma_{\s \text{out}}/\sigma_{\s \text{in}}$. It is seen from Eq.(\ref{15}) that the factor ${\cal M}$ depends strongly on the relation between the radius of the sphere and the spin-diffusion length. For $a\ll \lambda$
the last term in Eq. (\ref{15}) is $1$, while for $\lambda \ll a$ it is big and equal to $a/\lambda$.  In the latter case Eq. (\ref{14}) yields  ${\cal P}_{\s c}\propto \lambda a^2$. This dependence has a simple interpretation. Namely, for $\lambda \ll a$ the induced dipole is generated only inside a spherical layer of a thickness $\sim \lambda$ near the surface of the sphere, see Fig. \ref{Fig2}.

Description of a direct spin Hall effect for a sphere is completely similar to the case of the inverse spin Hall effect considered above. A charge current, $i_{\s c}$, along the $y$ direction generates a spin dipole moment, ${\cal P}_{\s s}$, in the $z$-direction. Analytical expression for   ${\cal P}_{\s s}$ is similar to Eq. (\ref{14})
\begin{equation}
\label{16}
{\cal P}_{\s s}=\frac{3 \sigma_{\s \text{in}} a^3\gamma  }{e (\sigma_{\s \text{in}}+2\sigma_{\s \text{out}})D_{\s \text{in}}{\cal M}} i_{\s c}.
\end{equation}

 \subsection{Finite density of grains}

\begin{figure}
\includegraphics[width=70mm]{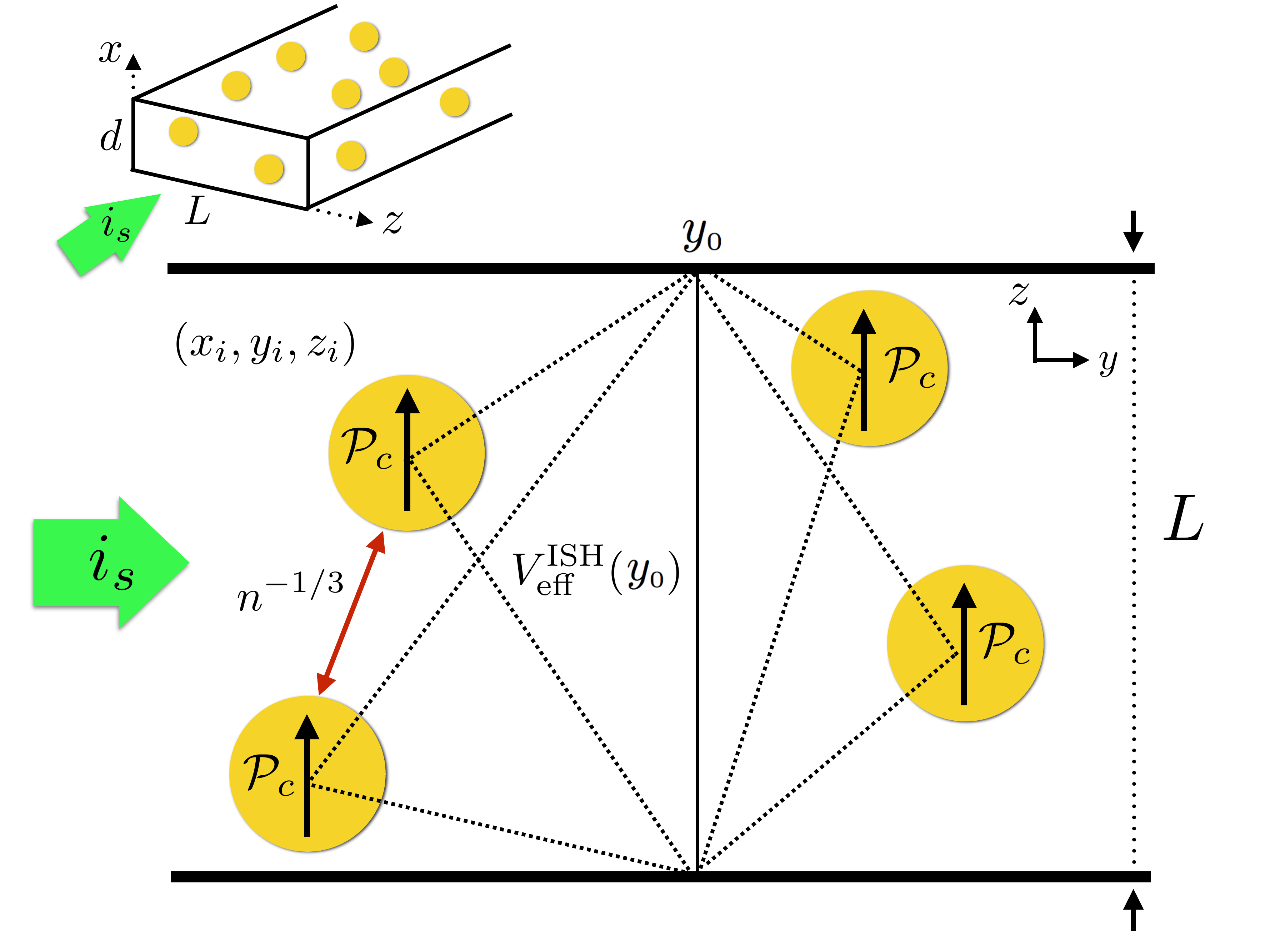}
\caption{(Color online)  The effective spin Hall voltage is the sum of contributions from individual SO-induced dipoles.  With density of granules, $n$, the typical distance between the neighbors is $n^{-1/3}$. It is much bigger than the radius, $a$, but much smaller than the sample width, $L$, which allows to replace the sum by the integral Eq. (\ref{18}).}
\label{Fig3}
\end{figure}

Consider a sample of a rectangular shape with a width, $L$, and thickness, $d$,  ($L\gg d$).
As the injected spin current flows through the cross section, the voltage builds up between
the edges $z=\pm L/2$. The easiest way to calculate this voltage is to sum the
contributions of individual dipoles. If a grain is located at a point with coordinates $(x_{\s i},y_{\s i},z_{\s i})$,
see Fig.  \ref{Fig3}, then the potential difference between the edges, created by an induced dipole
reads
\begin{align}
\label{17}
V(x_{\s i},y_{\s i},z_{\s i})=&\frac{(\frac{L}{2}-z_{\s i}){\cal P}_{\s c}}{[x_{\s i}^2+y_{\s i}^2+(\frac{L}{2}-z_{\s i})^2]^{\frac{3}{2}}} \nonumber\\
&-\Big(-\frac{(\frac{L}{2}+z_{\s i}){\cal P}_{\s c}}{[x_{\s i}^2+y_{\s i}^2+(\frac{L}{2}+z_{\s i})^2]^{\frac{3}{2}}}\Big),
\end{align}
where ${\cal P}_{\s c}$ is given by Eq. (\ref{14}). In calculating the effective inverse spin Hall  voltage the summation over dipoles is replaced by integration
\begin{equation}
\label{18}
V_{\s \text{eff}}^{\s \text{SH}}(y_{\s 0})=n\int_{-\frac{d}{2}}^{\frac{d}{2}}dx\int_{-y_{\s 0}}^{\infty}dy
\int_{-\frac{L}{2}}^{\frac{L}{2}}dz V(x,y,z),
\end{equation}
where $y_{\s 0}$ is the distance from the point at which voltage is measured to the point of spin-current injection. Naturally, the replacement of the sum by integral is justified when $nL^2d\gg 1$. The integration over $y$  is straightforward. Subsequent integral
over $z$ diverges logarithmically at  $z=L/2$ and $z=-L/2$. This divergence  should be cut off
at $(z\pm L/2)\sim d$. Then the integration over $x$ reduces to multiplication by $d$. The final
result reads
\begin{multline}
\label{19}
V_{\s \text{eff}}^{\s \text{SH}}(y_{\s 0})=2nd{\cal P}_{\s c}\\
\times \Biggl[\ln\Big(\frac{L}{d}\Big)+\ln\Big(\frac{2y_{\s 0}}{d}\Big)
-\ln\Biggl(\frac{\sqrt{\frac{L^2}{y^2_{\s 0}}+1}+\frac{L}{y_{\s 0}}+1}{\sqrt{\frac{L^2}{y^2_{\s 0}}+1}+
\frac{L}{y_{\s 0}}-1}\Biggr)\Biggr].
\end{multline}
At small distances from the injection point, $d\ll y_{\s 0}\ll L$, the first two terms in Eq. (\ref{19}) dominate. The second logarithm describes a gradual increase of $V_{\s \text{eff}}^{\s \text{SH}}$
with $y_{\s 0}$. At large distances,  $y_{\s 0}\gg L$, the second and the third logarithms combine into $\ln(L/d)$
leading to the result
\begin{align}
\label{20}
V_{\s \text{eff}}^{\s \text{SH}}&(\infty)=4nd{\cal P}_{\s c}\ln\Big(\frac{L}{d}\Big) \nonumber \\
=&-\frac{24e(na^3)d\ln(\frac{L}{d})\gamma }{(\sigma_{\s \text{in}}+2\sigma_{\s \text{out}} )\Big[\frac{2D_{\s \text{out}}}{D_{\s \text{in}}}+\frac{a i'_1(\frac{a}{\lambda})}{\lambda i_1(\frac{a}{\lambda})}\Big]}i_{\s s}.
\end{align}
Note, that for highly conducting grain, both factors in denominator do not depend on the characteristics,  $\sigma_{\s \text{out}}$ and  $D_{\s \text{out}}$, of the matrix. In this domain
$V_{\s \text{eff}}^{\s \text{SH}}$ depends strongly on the relation between $\lambda$ and $a$.
Overall, Eq. (\ref{19}) describes the growth and subsequent saturation of the inverse spin Hall
voltage.
%It is convenient to cast Eq. (\ref{20}) as the relation between the effective spin-Hall angle, $\theta_{\s \text{eff}}^{\s \text{SH}}$, of the mixture and the spin-Hall angle, $\theta^{\s \text{SH}}$, of the material of the grain.
%\begin{equation}
%\label{20a}
%\theta_{\s \text{eff}}^{\s \text{SH}}=\frac{12(na^3)\ln\Big(\frac{L}{d}\Big)}{\frac{2D_{\s \text{out}}}{D_{\s \text{in}}}+\frac{a i'_1(\frac{a}{\lambda})}{\lambda i_1(\frac{a}{\lambda})}}\theta^{\s \text{SH}},
%\end{equation}
%where we assumed $\sigma_{\s \text{in}}\gg \sigma_{\s \text{in}}$. Essentially, the
%proportionality between $\theta_{\s \text{eff}}^{\s \text{SH}}$ and $\theta^{\s \text{SH}}$ is determined by a ``volume factor", $na^3$.
%Note, however, that $\theta^{\s \text{SH}}$ is the characteristics of a homogeneous SO-film, only as long as the film thickness, $w$, is much smaller than $\lambda$. For $w \gg \lambda$, $\theta^{\s \text{SH}}$ falls off as $\lambda/w$. At the same time, the decay of $\theta_{\s \text{eff}}^{\s \text{SH}}$ with $y_{\s 0}$ sets in only when $y_{\s 0}$ exceeds the {\em effective} spin diffusion
%length of the mixture. This length is much bigger than $\lambda$. It is calculated in Section IV.

\section{Magnetic-field dependence}
\begin{figure}
\includegraphics[width=70mm]{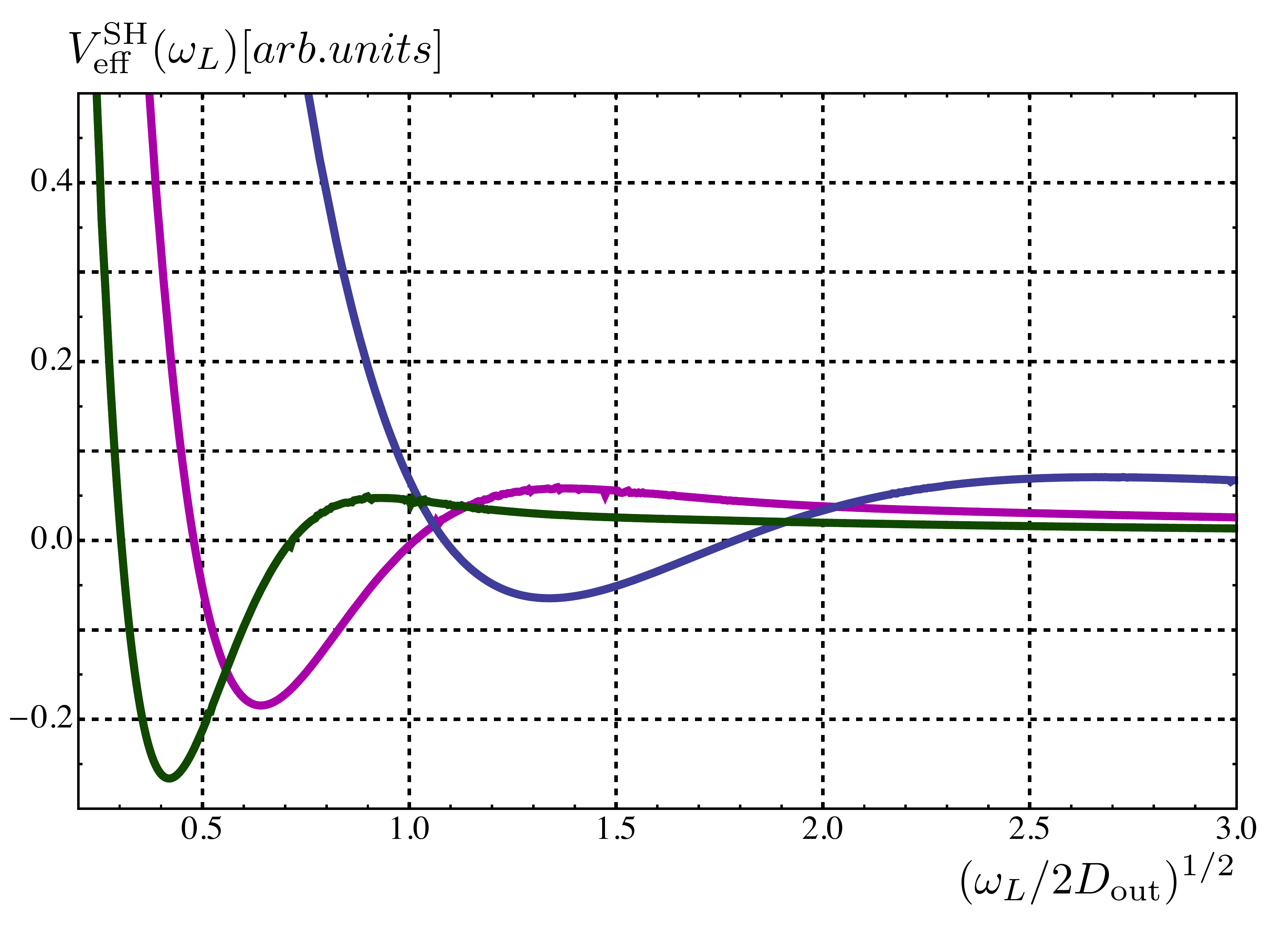}
\caption{(Color online) Dependence of the effective ISHE voltage on a longitudinal magnetic field, $\omega_{\s L}$, is plotted from Eq. (\ref{23}) for three different positions.
$y_{\s 0}$, in the units of $(dL)^{1/2}$, along the sample. Blue, violet, and green curves correspond to $y_{\s 0}=2$,
$y_{\s 0}=4$, and $y_{\s 0}=6$ , respectively.}
\label{Fig4}
\end{figure}
The behavior of $V_{\s \text{eff}}^{\s \text{SH}}$ with position, $y_{\s 0}$, becomes nontrivial in the
presence of magnetic field directed along the $y$-axis,
a somewhat similar effect was pointed out in Ref. \onlinecite{DyakonovMagnetoresistance}.
%see Fig. \ref{Fig4}.
If the magnetic field
is weak, so that the Larmour frequency, $\omega_{\s L}$, is much smaller that $\tau_{\s s}^{-1}$ and much smaller than $D_{\s \text{out}}/a^2$, which is the inverse diffusion time through the grain, then
the effect of magnetic field on generation of electric dipole can be neglected. Instead, the field
affects only the polarization in the spin current incident on the grain. This allows one to use the result
Eq. (\ref{14}) in calculation the $\omega_{\s L}$-dependence of $V_{\s \text{eff}}^{\s \text{SH}}$.

Outside  the grains, the polarization components, $P_x$ and $P_z$, satisfy the system of equations: $D_{\s \text{out}}\frac{d^2P_x}{dy^2}+\omega_{\s L}P_z=0$ and $D_{\s \text{out}}\frac{d^2P_z}{dy^2}-\omega_{\s L}P_x=0$. Assuming that at the point of injection the
polarization was along $x$, we find
\begin{align}
\label{21}
P_x(y)=&P_x(0)\cos\Big[\Big(\frac{\omega_{\s L}}{2D_{\s \text{out}}}\Big)^{\frac{1}{2}}y\Big] \exp\Big[-\Big(\frac{\omega_{\s L}}{2D_{\s \text{out}}}\Big)^{\frac{1}{2}}y\Big], \nonumber \\
P_z(y)=&P_x(0)\sin\Big[\Big(\frac{\omega_{\s L}}{2D_{\s \text{out}}}\Big)^{\frac{1}{2}}y\Big] \exp\Big[-\Big(\frac{\omega_{\s L}}{2D_{\s \text{out}}}\Big)^{\frac{1}{2}}y\Big].
\end{align}
Suppose that a grain is positioned at $y=y_{\s 0}$. Then the induced dipole moment will be a vector orthogonal to polarization with components
\begin{equation}
\label{22}
{\cal P}_z(y_{\s 0})=\left(\frac{P_x(y_{\s 0})}{P_x(0)}\right){\cal P}_{\s c},~~~~{\cal P}_x(y_{\s 0})=-\left(\frac{P_z(y_{\s 0})}{P_x(0)}\right){\cal P}_{\s c},
\end{equation}
where ${\cal P}_{\s c}$, given by Eq. (\ref{14}), is proportional to the magnitude of the spin
current, $i_{\s s}$, which does not change in the presence of magnetic field.
To proceed further, we notice that only ${\cal P}_z$-component of the induced dipole moment contributes to the buildup of  $V_{\s \text{eff}}^{\s \text{SH}}$ and should be substituted into Eq. (\ref{17}) instead of ${\cal P}_{\s c}$. We first perform integration over $z$ and $x$. The remaining integral over $y$ takes the form
\begin{widetext}
\begin{equation}
\label{23}
V_{\s \text{eff}}^{\s \text{SH}}(y_{\s 0},\omega_{\s L})=2nd{\cal P}_{\s c} \int_{0}^{\infty}dy \Biggl[\frac{1}{\sqrt{(y-y_{\s 0})^2+d^2}} -\frac{1}{\sqrt{(y-y_{\s 0})^2+L^2}} \Biggr]
 \cos\Big[\Big(\frac{\omega_{\s L}}{2D_{\s \text{out}}}\Big)^{\frac{1}{2}}y\Big] \exp\Big[-\Big(\frac{\omega_{\s L}}{2D_{\s \text{out}}}\Big)^{\frac{1}{2}}y\Big].
\end{equation}
\end{widetext}
For $\omega_{\s L}=0$ Eq. (\ref{23}) reproduces the limiting cases of Eq. (\ref{19}).
With characteristic distance $y_0$, being $\sim L$, we conclude that characteristic magnetic
field is
\begin{equation}
\label{24}
{\tilde \omega}_{\s L}=\frac{1}{T}=\frac{D_{\s \text{out}}}{L^2},
\end{equation}
which is a natural scale at which the diffusion time through a square with a side $L$ is
equal to the Larmour period. Simple asymptotic expressions for $V_{\s \text{eff}}^{\s \text{SH}}$
can be obtained in the domain $\omega_{\s L}\gg{\tilde \omega}_{\s L}$, when the second term in the integrand can be neglected:

(i) $d\ll y_{\s 0} \ll \Big(D_{\s \text{out}}/\omega_{\s L}\Big)^{1/2}$. In this limit, the log-divergence at large $y$ is cut off at $y\sim \left(D_{\s \text{out}}/\omega_{\s L}\right)^{1/2}$,
and we get
\begin{equation}
\label{25}
V_{\s \text{eff}}^{\s \text{SH}}(y_{\s 0},\omega_{\s L})=2nd{\cal P}_{\s c}\ln\Big(\sqrt{\frac{D_{\s \text{out}}}{\omega_{\s L}}}\frac{y_{\s 0}}{d^2}\Big)
\end{equation}

(ii) $y_{\s 0} \gg \Big(D_{\s \text{out}}/\omega_{\s L}\Big)^{1/2}$. We can now neglect $y$ compared
to $y_{\s 0}$ in the square brackets. Then the integration can be easily performed yielding
\begin{equation}
\label{26}
V_{\s \text{eff}}^{\s \text{SH}}(y_{\s 0},\omega_{\s L})=nd{\cal P}_{\s c}\Big(\frac{D_{\s \text{out}}}{\omega_{\s L}y_{\s 0}^2}\Big)^{\frac{1}{2}}.
\end{equation}
The asymptotes Eq. (\ref{25}), (\ref{26}) do not cover the entire domain of $\omega_{\s L}$.
At the crossover field $\omega_{\s L}\sim D_{\s \text{out}}/y_{\s 0}^2$. Eq. (\ref{25}) exceeds
Eq. (\ref{26}) by a large factor $\sim \ln(y_{\s 0}/d)$. As the magnetic-field dependence of
voltage is plotted numerically, see Fig.  \ref{Fig4}, it appears that in the intermediated domain
the ISHE voltage exhibits two sign reversals.  This means that the oscillations in Eq. (\ref{21})
do not average out completely after integration over the positions of the spheres.

\section{Effective spin diffusion length}
 \begin{figure}
\includegraphics[width=70mm]{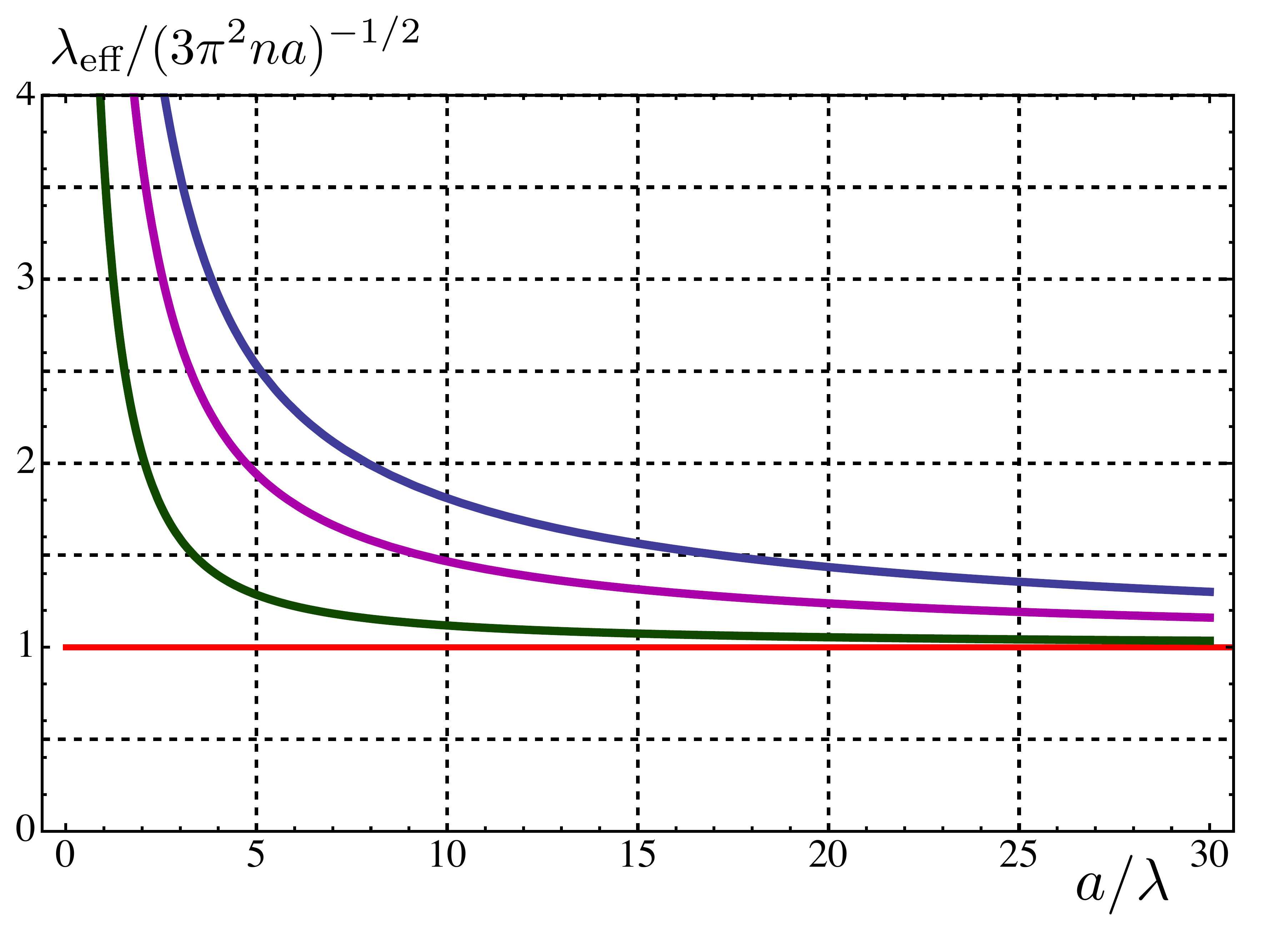}
\caption{(Color online) The effective spin diffusion length in the units $(3\pi^2na)^{-1/2}$ is plotted from Eq. (\ref{34}) for ratios
$D_{\s \text{out}}/ D_{\s \text{in}}$:  $D_{\s \text{out}}/ D_{\s \text{in}}=1$ (green),
$D_{\s \text{out}}/ D_{\s \text{in}}=5$ (purple),  and $D_{\s \text{out}}/ D_{\s \text{in}}=10$ (blue). Note the  saturation of $\lambda_{\s \text{eff}}$ at small $\lambda$  .}
\label{Fig5}
\end{figure}
There are two reasons why the effective spin-diffusion length of the mixture exceeds $\lambda$.
The first reason is obvious: the grains are sparse and there is no spin relaxation in between the grains. The second reason is much more subtle and becomes important when $\lambda$ is much smaller than the grain radius. Namely, the rate of the spin relaxation at the grain surface is suppressed. Formally, this suppression, illustrated in Fig. \ref{Fig2}, follows from the behavior of polarization inside the grain

\begin{align}
\label{27}
{\bm P}_{\s \text{in}}({\bm r})=&-\frac{3a}{D_{\s \text{in}}{\cal M}}\Big( \frac{i_1( \frac{r}{\lambda})}{i_1 (\frac{a}{\lambda})}\Big)i_{\s s}\sin\theta \hspace{0.5mm}{\bm e}_{\s \phi}\nonumber \\
=&-\frac{3ai_1 (\frac{r}{\lambda})}{2D_{\s \text{out}}i_1 (\frac{a}{\lambda})+D_{\s \text{in}}\frac{a}{\lambda}i'_1 (\frac{a}{\lambda})}i_{\s s}\sin\theta \hspace{0.5mm}{\bm e}_{\s \phi}.
\end{align}
It is seen from Fig. \ref{Fig2} that, for $\lambda =a/12$, the radial distribution of
${\bm P}_{\s \text{in}}({\bm r})$ not only falls off rapidly from the surface towards the
center, but its value at the surface is small. Physical origin of this smallness is elucidated in the Appendix.

While our goal is to find $\lambda_{\s \text{eff}}$, in order not to deal with boundaries we
first calculate the effective spin relaxation time of the mixture.
Spin relaxation takes place only inside the spheres.
If at time $t=0$ the polarization inside the sphere is distributed according to Eq. (\ref{27}),
then the rate of decay of this polarization is given by the integral over the volume of the sphere

\begin{equation}
\label{28}
R=\frac{1}{\tau_{\s s}}\int d\Omega \Big[{\bm P}_{\s \text{in}}({\bm r})\Big]_{\s \phi}.
\end{equation}
Using the explicit form, $i_1(x)=\left(x\cosh(x)-\sinh(x)\right)/x^2$, of the modified spherical Bessel function, the integral can be evaluated, and the result can be cast in the form

\begin{equation}
\label{29}
R=\frac{3\pi^2a^4i_{\s s}}{D_{\s \text{in}}\tau_{\s s}}{\mathrm{F}}\Big(\frac{a}{\lambda}\Big),
\end{equation}
where the dimensionless function ${\mathrm{F}}(x)$ is defined as
\begin{equation}
\label{30}
{\mathrm{F}}(x)=\frac{x\sinh(x)-2\cosh(x)+2}{2(\frac{D_{\s \text{out}}}{D_{\s \text{in}}}-1)[x\cosh(x)-\sinh(x)]x+x^3\sinh(x)}.
\end{equation}
The result Eq. (\ref{29}) can be also expressed through the polarization outside the sphere by
replacing $i_{\s s}$ by $P_{\s \text{out}}D_{\s \text{out}}/a$, see Eq. (\ref{8}). One has
\begin{equation}
\label{31}
R=\frac{3\pi^2a^3P_{\s \text{out}}D_{\s \text{out}}}{D_{\s \text{in}}\tau_{\s s}}{\mathrm{F}}\Big(\frac{a}{\lambda}\Big).
\end{equation}

In the absence of spin current, the spin relaxation inside the spheres causes the time decay of the spin polarization in the medium between the spheres. This is because diffusing carriers eventually
``hit" a sphere.
Consider an interval $(y_{\s 0}-\frac{\delta y}{2}, y_{\s 0}+\frac{\delta y}{2})$, and assume that there are hard walls at the ends, so electrons do not flow in or out. Then the initial net polarization, $P_{\s \text{out}}(y_{\s 0})\delta y$, inside the interval
will decay with some effective rate $\tau_{\s \text{eff}}^{-1}$.
To find this rate, we substitute the  the two-dimensional density of spheres in the interval, $n\delta y$, into the balance equation
\begin{equation}
\label{32}
\frac{P_{\s \text{out}}(y_{\s 0})\delta y}{\tau_{\s \text{eff}}}=n\delta y R.
\end{equation}
Substituting Eq. (\ref{31}) into Eq. (\ref{32}), we readily find
\begin{equation}
\label{33}
\tau_{\s \text{eff}}=\frac{D_{\s \text{in}}\tau_{\s s}}{3\pi^2na^3D_{\s \text{out}}{\mathrm{F}}(\frac{a}{\lambda})}.
\end{equation}
Note that the product in the numerator is equal to $\lambda^2$.  We can now use the expression for the effective relaxation time to find the effective spin-diffusion length
\begin{equation}
\label{34}
\lambda_{\s \text{eff}}=\sqrt{D_{\s \text{out}}\tau_{\s \text{eff}}}=
\frac{\lambda}{\big[3\pi^2na^3{\mathrm{F}}(\frac{a}{\lambda})\big]^{1/2}}.
\end{equation}

Let us trace the decrease of $\lambda_{\s \text{eff}}$ as the spin-diffusion rate inside the sphere gradually decreases. For $\lambda \gg a$, the function ${\mathrm{F}}(x)$ can be replaced by
${\mathrm{F}}(0)=\frac{D_{\s \text{in}}}{4( 2D_{\s \text{out}}+D_{\s \text{in}})}$. Thus the enhancement of the spin-diffusion length due to patterning the SO material into granules is $\sim (na^3)^{-1/2}$. In the opposite limit $\lambda \ll a$ we have
${\mathrm{F}}(x) \approx 1/x^2$. This leads to the unexpected conclusion that in this limit $\lambda_{\s \text{eff}}$ {\em saturates} at the value $\sim \frac{1}{(na)^{1/2}}$. The origin of this saturation is suppression of polarization at the surface, the effect discussed above and further elaborated on in Appendix.

\section{Discussion}

({\em i}). The two main results of the present paper are  Eqs. (\ref{19}), (\ref{20}) and Eq. (\ref{34})
for the effective inverse spin Hall voltage and the effective spin diffusion length
of the mixture.
%Overall, Eq. (\ref{19}) describes the growth and subsequent saturation of the inverse spin Hall
%voltage.
It is convenient to cast Eq. (\ref{20}) in the form of the relation between the effective spin-Hall angle, $\theta_{\s \text{eff}}^{\s \text{SH}}$, of the mixture and the spin-Hall angle, $\theta^{\s \text{SH}}$, of the material of the grain. The spin Hall angle is defined as the proportionality coefficient between the charge and spin current densities,
more precisely, $j_{\s c} =\theta^{\s \text{SH}}(2e/\hbar)j_{\s s}$. Then Eq. (\ref{20}) takes the form

\begin{equation}
\label{20a}
\theta_{\s \text{eff}}^{\s \text{SH}}=\frac{12(na^3)\ln\Big(\frac{L}{d}\Big)}{\frac{2D_{\s \text{out}}}{D_{\s \text{in}}}+\frac{a i'_1(\frac{a}{\lambda})}{\lambda i_1(\frac{a}{\lambda})}}\theta^{\s \text{SH}},
\end{equation}
where we assumed $\sigma_{\s \text{in}}\gg \sigma_{\s \text{out}}$. Essentially, the
proportionality between $\theta_{\s \text{eff}}^{\s \text{SH}}$ and $\theta^{\s \text{SH}}$ is determined by a ``volume factor", $na^3$.
Note, however, that $\theta^{\s \text{SH}}$ is the characteristics of a homogeneous SO-film, only as long as the film thickness, $w$, is much smaller than $\lambda$. For $w \gg \lambda$, $\theta^{\s \text{SH}}$ falls off as $\lambda/w$. At the same time, the decay of $\theta_{\s \text{eff}}^{\s \text{SH}}$ with $y_{\s 0}$ sets in only when $y_{\s 0}$ exceeds the {\em effective} spin diffusion
length of the mixture. This length is much bigger than $\lambda$, as it was shown in  Section IV.

({\em ii}) Note that, strictly speaking, Eq. (\ref{34})
describes $\lambda_{\s \text{eff}}$ only within a numerical factor. This factor was lost  as we replaced $i_{\s s}$ by $P_{\s \text{out}}D_{\s \text{out}}/a$, assuming that the
first term in Eq. (\ref{8}) dominates. In fact, precisely at $r=a$, the two terms almost cancel
each other. Indeed, substituting the expression for $\chi_{\s s}$ into Eq. (\ref{8}), we can cast it in the form
\begin{equation}
\label{35}
{\bm P}_{\s \text{out}}=-\frac{i_{\s s}}{D_{\s \text{out}}}\Biggl(r-\frac{a^3}{r^2}+\frac{3a^3}{\Big[2+\frac{aD_{\s \text{in}} i'_1(\frac{a}{\lambda})}{\lambda D_{\s \text{out}} i_1(\frac{a}{\lambda})}\Big]r^2}\Biggr)\sin\theta \hspace{0.5mm}{\bm e_{\s \phi}}.
\end{equation}
In the limit $\lambda \ll a$  and $r=a$,  the expression  in the brackets  is equal to
$3\lambda D_{\s \text{out}}/D_{\s \text{in}}$, and thus, is much smaller than $a$. However, for bigger $r\sim a$ the compensation of the first two terms does not take place, and the relation $i_{\s s} \sim P_{\s \text{out}}D_{\s \text{out}}/a$ holds.

The suppression of $P_{\s \text{out}}$ near the surface of the sphere, expressed by
Eq. (\ref{35}), is the reason why $\lambda_{\s \text{eff}}$ saturates when $\lambda \rightarrow 0$, see Fig. \ref{Fig5}.
Loosely speaking, strong relaxation ``repels" the spins from the boundary, which, in turn, slows down the effective
relaxation. The above physics is quite general. To illustrate it, in the Appendix we consider a model example of diffusion of particles in the presence of an absorbing trap and demonstrate that, with increasing the absorption rate, the concentration of particles vanishes at the position of the trap.

({\em iii}). It is instructive to compare our result Eq. (\ref{20}) with the
 expression for the perturbation of spin current flowing in a normal metal around a ferromagnetic sphere\cite{Roundy}. Rather that the SO coupling in our case, the difference of spin-up and spin-down carriers in Ref. \onlinecite{Roundy} is caused by the difference of their conductivities inside the ferromagnet. As a result, the induced dipole moment in our case is normal to the spin current, while the induced ``spin dipole moment"\cite{Roundy}    is along the spin current. Other than that, the two expressions resemble each other.
There is, however, an important difference. If the conductivity
of the ferromagnetic sphere\cite{Roundy} is much higher than the conductivity of the surrounding normal medium, then the perturbation of the spin current is suppressed (resistance mismatch). On the contrary, for the inverse spin Hall effect,  the bigger is the ratio $D_{\s \text{in}}/D_{\s \text{out}}$, the stronger is the modification of the
spin current outside the sphere.

 ({\em iiii}). For a quantitative example of the effect of granularity on the effective
 parameters of the mixture, assume that the density of the SO granules is $na^3=10^{-2}$, while the spin diffusion length in the material of the granule is $\lambda =0.2a$. Compared to the geometry in Fig.  \ref{Fig1} with no granularity we ``lose" 100 times in the inverse spin Hall voltage. At the same time, we gain in $\lambda_{\s \text{eff}}$. Substituting $\lambda =0.2a$ into Eq. (\ref{34}), and assuming $D_{\s \text{in}} \gg D_{\s \text{out}}$, we find $\lambda_{\s \text{eff}}=10\lambda$.

\section{Concluding remarks}

1.	For experimentally verifying our theoretical results, composites of SO and no SO materials can be prepared using a variety of widely available fabrication techniques. For example, in Ref. \onlinecite{WW}, authors used a pulsed laser deposition technique to prepare a composite comprising of gold nanoclusters embedded in ZnO matrix.
In Ref. \onlinecite{XX}, a self-assembly approach was used to fabricate a composite comprising of nickel nanoclusters embedded in amorphous Al$_2$O$_3$ matix. In Ref. \onlinecite{MagnonicCrystals}, a nanofabrication approach was employed to prepare a magnonic crystal comprising of cobalt nanodots embedded in a permalloy film. Similar approaches can be used to prepare the desired composite structures of SO and no SO materials, say Pt or Au nanodots (with large $\theta^{\s \text{SH}}$) embedded in films of low $\theta^{\s \text{SH}}$ materials (such as copper, molybdenum or even semiconductors like silicon).

%
%1. One of approaches  to make a composite of SO and no SO materials is to employ  an
%experimental technique for fabrication magnonic crystals\cite{MagnonicCrystals}.
% % An experimental technique\cite{MagnonicCrystals} for fabrication magnonic crystal can probably be used to realize a system of SO spheres in a matrix with no SO.
% An objective of this technique is to
%fabricate a mixture of two ferromagnets, like permalloy and cobalt, in which the spectrum of spin
%waves possesses a gap. If platinum with large $\theta^{\s \text{SH}}$ could  replace cobalt, while copper (or molybdenum) with small $\theta^{\s \text{SH}}$ could replace permalloy, the resulting ``crystal" would represent a desired mixture.
%There are other approaches\cite{MixtureCo-ZnO,ZnONano-GranularFilms,CoInPNanocomposites,NanoparticlesInSilicon} like sputtering or elctrodeposition
%to fabricate the composites of materials with very different magnetic properties in the form of
%the films.

 2. From device point-of-view, an obvious way to enhance the spin diffusion length would be by creating a 1D structure of alternating SO and no SO layers
To achieve this, however, the thickness of the SO-layer should be smaller
than $\lambda$. Conversely, in granular system the enhancement takes place when $\lambda \ll a$. This is because the spin current can flow {\em around} the spheres.

3. In numerous spin-pumping experiments, see e.g. Refs. \onlinecite {Pt1,Pt2,Pt3,Pt4,Pt5,Pt6,Bader,Gold,Exotic,GaAs,silicon1,silicon2,germanium,graphene}, the measured
quantity, $V^{\s \text{SH}}$, is proportional either to $\theta^{\s \text{SH}}\lambda$, when the thickness of non-magnetic material is much bigger than $\lambda$, or simply to  $\theta^{\s \text{SH}}$ in the opposite limit. A comprehensive list of experimental values of $\theta^{\s \text{SH}}$ and $\lambda$ for a number of heavy metals can be found in  Ref.~\onlinecite{LambdaInMANYmetals}. This list indicates that, while, {\em separately},
$\theta^{\s \text{SH}}$, and $\lambda$ vary within wide ranges, the range of change of
their product is much narrower, see also Ref. \onlinecite{InALLOY}.
%
%In a specific case of a semiconductor ZnO the inverse spin Hall effect was studied both
%in pumping experiment\cite{ZnO} and directly by measuring the nonlocal voltage\cite{ZnOTiwari}.
%In both measurements the value $\theta_{\s \text{SH}}$ was found to be anomalously big, compared e.g. to Si\cite{silicon1,silicon2}.

Overall, there is still experimental ambiguity in extracting the intrinsic SO parameters of materials
from the experiment. In this regard, granularity can offer a help, by bringing a new spatial scale,
the radius of the grain, $a$. As shown in Fig. \ref{Fig5}, the value $\lambda_{\s \text{eff}}$ depends
very strongly on the relation between $\lambda$ and $a$.

4.	In a specific case of a semiconductor ZnO the inverse spin Hall effect was studied both in pumping experiment\cite{ZnO} and directly by measuring the nonlocal voltage\cite{ZnOTiwari}. In both measurements the value $\theta^{\s \text{SH}}$ was found to be anomalously big, compared e.g. to Si\cite{silicon1,silicon2} . It has recently been shown\cite{ZZ} that the value $\theta^{\s \text{SH}}$ in ZnO can be tuned very sensitively by changing the oxygen ambient under which it is grown\cite{YY}. Films prepared under high oxygen rich environment showed a large value for $\theta^{\s \text{SH}}$ $(\sim 0.1)$, while the films prepared under low oxygen ambient showed an order of magnitude lower value of $\theta^{\s \text{SH}}$.

%In Ref.\onlinecite{InALLOY} $\theta_{\s \text{SH}}=0.1$, $\lambda=2nm$

%The values of  $\theta_{\s \text{SH}}$ from Ref. \onlinecite{LambdaInMANYmetals} and their correlation with spin diffusion lengths:
%
%1. Pt  $\theta_{\s \text{SH}}=0.1$  $\lambda=7.3nm$  Product 0.73
%
%2. Ta $\theta_{\s \text{SH}}=-0.07$  $\lambda=1.9nm$  Product 0.133
%
%3. W $\theta_{\s \text{SH}}=-0.14$  $\lambda=2.1nm$   Product 0.294
%
%4. Au $\theta_{\s \text{SH}}=-0.084$  $\lambda=60nm$   Product 5.4
%
%5. Ag $\theta_{\s \text{SH}}=-0.0068$  $\lambda=700nm$  Product 4.76
%
%
%
%
%
%6. Cu $\theta_{\s \text{SH}}=-0.0032$  $\lambda=500nm$   Product 1.6
%
%According to Ref. \onlinecite{ZnOTiwari}
%
%7. ZnO $\theta_{\s \text{SH}}=-0.01651$
%
%According to Ref. \onlinecite{ZnO}
%
%8. Si $\theta_{\s \text{SH}}=-0.0001$  $\lambda=200-300nm$ Product 0.03
%
%9. Py $\theta_{\s \text{SH}}=0.005$  $\lambda=3.3nm$  Product 0.0165
%
%10. Product $\theta_{\s \text{SH}}\lambda$ in ZnO is the same as in Py
%
%%The lattice constant of Pt is 3.92 angstrom, while in Cu 3.61 angstrom, and in Ag 4.08 angstrom. In GaAs 5.65 angstrom, InAs 6.0584 angstrom.

\section{Acknowledgements}
The authors are grateful to E. G. Mishchenko  for a useful discussion. The work was supported by NSF through MRSEC DMR-1121252.

\appendix
\section{}
 \begin{figure}
\includegraphics[width=82mm]{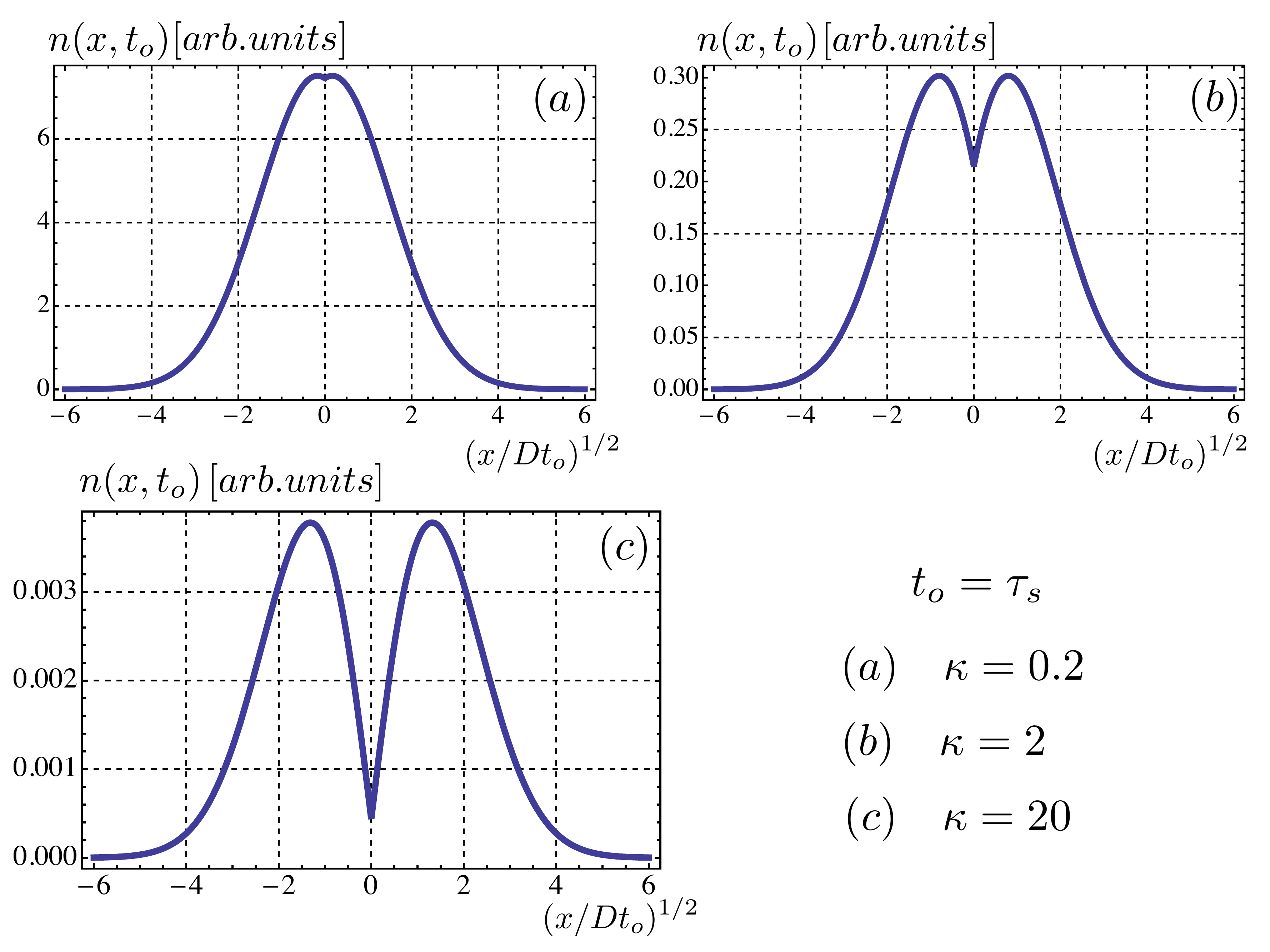}
\caption{Diffusive spreading of the initial particle distribution $n(x,0)=\delta(x)$ in the presence of an absorbing trap located  at $x=0$ is described by Eq. (\ref{A10}). Shown is $n(x,t)$ at a fixed time, $t_{\s 0}$, for different absorption efficiencies, $\kappa$,
Eq. (\ref{A13}). The more absorbing the trap is, the deeper is the dip at the origin, and the slower is the decay of the net number of particles at long times, as follows from
Eq. (\ref{A12}).    }
\label{Fig6}
\end{figure}

Consider a diffusion in one dimension.
If at time $t=0$ the  distribution of particles    is a $\delta$-peak, i.e. $n(x,0)=\delta(x)$, then it spreads with time as
\begin{equation}
\label{A1}
n(x,t)=  \frac{1}{2\sqrt{\pi D t}} \exp\Big(-\frac{x^2}{4Dt}\Big)
\end{equation}
where $D$ is the diffusion coefficient.

Suppose now,  that an absorbing trap is placed at the coordinate origin. Then the spreading is governed by the equation
\begin{equation}
\label{A2}
-\frac{\partial n}{\partial t} =-D \frac{\partial^2 n}{\partial x^2} +\frac{a\delta(x)n}{\tau_{\s s}}
\end{equation}
where $a$ is the size of the trap, and $\tau_{\s s}$ is the absorption rate.
Then the time-dependent concentration, $n(x,t)$, can be expressed through the
Green function of Eq.~(\ref{A2})
\begin{equation}
\label{A3}
n(x,t)=\int dx' G(x,x',t) n(x',0)
\end{equation}
It is convenient to present the Green function in terms of eigenfunctions, $\psi_{k}$ of Eq. (\ref{A2}),
which satisfy the Sch{\"o}dinger-like equation
\begin{equation}
\label{A4}
-D\frac{\partial^2 \psi_k}{\partial x^2} +\frac{a\delta(x)}{\tau_{\s s}}\psi_k=k^2\psi_k.
\end{equation}
Then the expression for G(x,x',t) reads
\begin{equation}
\label{A5}
G(x,x',t)=\sum_{k}\psi_k(x)\psi_k(x')\exp(-Dk^2t).
\end{equation}
The second term in Eq. (\ref{A4}) plays the role of delta-potential barrier, and causes the discontinuity of the derivative of $\psi_k$
\begin{equation}
\label{A6}
\frac{\partial \psi_k}{\partial x}\Big\vert_{x=0^+}-\frac{\partial \psi_k}{\partial x}\Big\vert_{x=0^-}=\frac{a}{D\tau_{\s s}}\psi_k(0).
\end{equation}
The normalized solutions, $\psi_k(x)$, which satisfy Eq. (\ref{A4}) have the form
\begin{equation}
\label{A7}
\psi_k(x)=\frac{1}{\pi^{1/2}}\cos(k|x| +\varphi_k),
\end{equation}
where the phase $\varphi_k$ is found from the condition
\begin{equation}
\label{A8}
\tan\varphi_k=-\frac{a}{2D\tau_{\s s}k},
\end{equation}
imposed by Eq. (\ref{A6}).

Upon substituting Eq. (\ref{A5}) into Eq. (\ref{A3}) and using the initial condition,
we arrive at the final result

\begin{equation}
\label{A9}
n(x,t)=\frac{1}{\pi}\int_0^\infty d k \frac{\cos k|x|+\frac{a}{2D\tau_{\s s}k}\sin k|x|}{1+(\frac{a}{2D\tau_{\s s}k})^2}e^{-Dk^2t}.
\end{equation}
It is now convenient to introduce a dimensionless coordinate ${\tilde x}=x/(Dt)^{1/2}$ and the dimensionless time  dependent parameter ${\tilde t}=a^2t/4D\tau_{\s s}^2$. In new variables Eq. (\ref{A9}) assumes the form
\begin{equation}
\label{A10}
n(x,t)=\frac{a}{2\pi D\tau_{\s s}\sqrt{{\tilde t}}}\int_0^\infty d q\frac{q^2\cos|{\tilde x}|q+\sqrt{{\tilde t}}q\sin|{\tilde x}|q }{q^2+{\tilde t}}e^{-q^2}.
\end{equation}
We see that the characteristics of the trap, $a$ and $\tau_{\s s}$, enter only into rescaling of time. In Fig.  \ref{Fig6} we plot Eq. (\ref{A10}) for four different ${\tilde t}$. It is seen from  Fig.  \ref{Fig6} that, with
time,  the density $n(0,t)$ at the origin develops a dip.  The smaller is $\tau_{\s s}$, i.e. the more absorbing
is the trap, the sharper is the dip. This conclusion also follows from the long-time asymptote
of $n(x,t)$, when we neglect $q^2$ in denominator compared to ${\tilde t}$. Then the integration yields
\begin{align}
\label{A11}
n(x,t)\Big\vert_{{\tilde t}\gg 1}&=\frac{a}{8\sqrt{\pi} D\tau_{\s s}{\tilde t}^{\frac{3}{2}}}\Big (\sqrt{{\tilde t}}|{\tilde x}|+1\Big)e^{-\frac{{\tilde x}^2}{4}} \nonumber \\
&=\frac{\sqrt{D}\tau_{\s s}^2}{\sqrt{\pi}a^2t^{\frac{3}{2}}}\Big (\frac{a|x|}{2D\tau_{\s s}}+1\Big)e^{-\frac{ x^2}{4Dt}}.
\end{align}
This asymptote indicates that the ratio of concentrations at half-width, $x\approx (Dt)^{1/2}$,
and at the origin is $\sim {\tilde t}^{1/2}$, i.e. the dip is deep.

The next question we ask ourselves is how the total number of particles $\int\hspace{-0.5mm} dx~ n(x,t)$ decreases with time. Upon integration Eq. (\ref{A10}), we get

%for  ${\tilde t} \ll 1$, we got
%\begin{equation}
%n(x,t)=\begin{cases}
% \frac{a}{4\sqrt{\pi} D\tau_{\s s}\sqrt{{\tilde t}}}  \Big( e^{-\frac{{\tilde x}^2}{4}}+\sqrt{{\tilde t}}|{\tilde x}|\Big) ,~~~|{\tilde x}| \ll1, \\
% \frac{a}{4\sqrt{\pi}  D\tau_{\s s}\sqrt{{\tilde t}}} \Big( e^{-\frac{{\tilde x}^2}{4}}+\sqrt{{\tilde t}}\Big) ,~~~~~~ |{\tilde x}|\gg1.
% \end{cases}
%\end{equation}
\begin{align}
\label{A12}
N(t)=&\int_{-\infty}^{\infty} dx ~n(x,t) \nonumber \\
=& \text{Erfc}({\tilde t}^{\frac{1}{2}})\exp({\tilde t})=
\begin{cases}
 1 -\frac{a}{\tau_{\s s}}\Big(\frac{t}{\pi D}\Big)^{\frac{1}{2}},~{\tilde t} \ll1, \\
\frac{2\tau_{\s s}}{a}\Big(\frac{\pi D}{t}\Big)^{\frac{1}{2}},~~{\tilde t}\gg1,
 \end{cases}
\end{align}
where $\text{Erfc}(s)$ is the complementary error function. It is seen from Eq. (\ref{A12})
that the change of the decay rate $\partial N/\partial t$ takes place at ${\tilde t}\sim 1$.
This change is caused by the development of the dip. Indeed, for ${\tilde t}\gg 1$ the decay rate
falls off with time as $t^{-3/2}$.

Overall, we conclude that the spreading of the particle density in the presence of a trap is
governed by a dimensionless parameter
\begin{equation}
\label{A13}
\kappa=\frac{a}{(D\tau_{\s s})^{1/2}},
\end{equation}
which is the dimensionless efficiency of absorption by the trap. If this efficiency is small, the spreading will proceed as in the absence of the trap  for most of the time, until the
concentration at $x=0$ becomes really small. Only then, $n(x,t)$ will develop a dip at $x=0$ and
the decay of the net number of particles will proceed even slower.
For large efficiency, the dip will developed early, namely at $t\sim \tau_{\s s}/\kappa^2\ll \tau_{\s s}$, after which time the
decay of $N(t)$ will be governed by
the value of $n(0,t)$ of the concentration at the dip.

Formation of a dip in our model problem puts into a general perspective the behavior of the effective spin diffusion length in the system of the SO grains. In the limit $\lambda \ll a$, see Eq. (\ref{34}), the value $\lambda_{\s \text{eff}}$ saturates because the  polarization near the boundary gets suppressed as a result of the development of a local minimum.

%{\bf Suppose that $\tau_{\s s}$ is negative, $\tau_{\s s}=-|\tau_{\s s}|$. This would result in multiplication of the initial number of particles near $x=0$. Remarkably, despite of diffusion, the particles
%will stay close to the point where they are multiply. The distribution $n(x,t)$ is given by}
%\begin{widetext}
%\begin{equation}
%\label{A14}
%n(x,t)=\frac{a}{ D|\tau_{\s s}|}\Big[\frac{1}{2\pi\sqrt{{\tilde t}}}\int_0^\infty d q\frac{q^2\cos|{\tilde x}|q-\sqrt{{\tilde t}}q\sin|{\tilde x}|q }{q^2+{\tilde t}}e^{-q^2}+e^{-{\tilde x}\sqrt{{\tilde t}}+{\tilde t}}\Big].
%\end{equation}
%\end{widetext}

\end{document}